

\input harvmac

%

\def\hat{\widehat}
\def\*{\star}
\def\({\left(}		
\def\){\right)}		
\def\[{\left[}		\def\BBL{\Bigl[}
\def\]{\right]}		\def\BBR{\Bigr]}

%
%
\def\frac#1#2{{#1 \over #2}}
\def\inv#1{{1 \over #1}}
\def\half{{1 \over 2}}
\def\d{\partial}

\def\vev#1{\langle #1 \rangle}
\def\ket#1{ | #1 \rangle}
\def\bra#1{ \langle #1 |}
\def\rvac{\hbox{$\vert 0\rangle$}}
\def\lvac{\hbox{$\langle 0 \vert $}}
\def\2pi{\hbox{$2\pi i$}}

\def\dsl{\raise.15ex\hbox{/}\kern-.57em\partial}
\def\Dsl{\,\raise.15ex\hbox{/}\mkern-.13.5mu D}
\def\oti{\otimes}
%
%
\def\th{\theta}		
		\def\Ga{\Gamma}

\def\al{\alpha}

\def\la{\lambda}	\def\La{\Lambda}
\def\de{\delta}		\def\De{\Delta}
		
\def\sig{\sigma}	
\def\vphi{\varphi}
%
%

%
\def\partk{\lbrace k \rbrace}
\def\partr{\lbrace r \rbrace}
\def\partkp{\lbrace k' \rbrace}
\def\sN{\sum_{i=1}^N}
\def\partki{\lbrace k_i \rbrace}
\def\vecw{(w_1,\ldots,w_N)}
\def\pref{\prod_{i<j}(w_i-w_j)^{1/2}}
\def\pphi{\(\prod_{i=1}^N \phi_i(w_i)\)}

\def\pphia{\(\prod_{i=1}^N \phi_i^{\al_i}(w_i)\)}
\def\pphip{\(\prod_{i=1}^N \phi_i^+(w_i)\)}
\def\Dj{\hat D_j}
\def\modn{\phi_n^{\pm} \rvac}
\def\partkN{k_1 \ge \ldots \ge k_N}
\def\veck{k_1,\ldots,k_N}
\def\dw{w_i \d_{w_i}}
\def\ketk{\ket{\partk}}


\Title{SPhT/94/039}{Spinons in Conformal Field Theory.}

\centerline{D. Bernard, V. Pasquier and D. Serban.}

\centerline{Service de Physique Th\'eorique,}
\centerline{CE Saclay, 91191 Gif-sur-Yvette, France}

\vskip 1.5 truecm
Abstract

We study the $su(2)$ conformal field theory in its spinon description,
adapted to the Yangian invariance. By evaluating the action of
the Yangian generators on the primary fields, we find a new
connection between this conformal field theory and the
Calogero-Sutherland model with $su(2)$ spin. We use this connection
to describe how the spinons are the quasi-particles spanning
the irreducible Yangian multiplet, and also to exhibit operators
creating the $N$-spinon highest weight vectors.

\Date{02/94}

\newsec{Introduction}

The spin chain with long range interaction
possesses remarkable properties \ref\Ha{F.D.M.Haldane,
Phys.Rev.Lett. 60 (1988) 635}\ \ref\Sha{B.S.Shastry, Phys.Rev.Lett.
60 (1988) 639}\ \ref\Hal{F.D.M.Haldane, Phys.Rev.Lett. 67 (1991)
937}~: its elementary excitations
can be described in terms of free particles, the spinons,
obyeing a fractional statistics. Most of the properties
of the spin chain survive in the conformal limit
\ref\nous{F.D.M.Haldane, Z.N.C.Ha, J.C.Talstra, D.Bernard and V.Pasquier,
Phys.Rev.Lett. 69 (1992) 2021}, which is described by the
$su(2)$ level one WZW model.

In particular, this gives an alternative description of the
spectrum of the $su(2)_{k=1}$ theory in terms of spinons.
This description is closely related to the new character
formula for the minimal models \ref\MC{S.Dasmahapatra,
R.Dedem, T.R.Klassen, B.M.McCoy, E.Melzer, Int.J.Mod.Phys.B
7 (1993) 3617 and references therein}.

In this paper we relate the $su(2)$ WZW model at level one
to the Calogero-Sutherland model with spins \ref\Po{A.P.Polychronakos,
Phys.Rev.Lett. 69 (1992) 703}\ \ref\Hi{K.Hikami and M.Wadati,
J.Phys.Soc.Jpn. 62 (1993)}\ \ref\Che{I.Cherednik, RIMS-742 preprint
(1990)}\ \ref\YB{D.Bernard, M.Gaudin, F.D.M.Haldane and V.Pasquier
J.Phys.A 26 (1993) 5219}, alternatively
called the dynamical model. Continuing
the work of \nous\ we show that there is a representation
of the Yangian algebra in the conformal field theory,
commuting with a set of hamiltonians. The first of these
hamiltonians is the generator $L_0$ of the Virasoro algebra
and the second and third  are natural extensions of the spin chain
Hamiltonians. In ref. \YB , it was shown that
the Calogero-Sutherland model with spin is also Yangian
invariant. The connection between the two models is
established by showing that the preceding operators of the
conformal field theory act on a product of primary
fields as the corresponding operators of the Calogero-Sutherland
model.

A relation between the $1/r^2$ spin chain and the Calogero-Sutherland
model was already known, and actually used to solve the spin
chain \Ha\ \Sha . The connection is established by constructing the spin
chain Hilbert space over the ferromagnetic vacuum~: it leads
to a coupling constant $\la=-2$ for the Calogero-Sutherland
model, in the normalisation of ref. \YB. Here, the relation
between the two models is different. The Hilbert space is
constructed over the conformal
vacuum which is the antiferromagnetic vacuum of the spin chain~:
it leads to a coupling constant $\la=-1/2$. This is an
illustration of the ``mirror" duality $\la\to 1/\la$ of
the Calogero-Sutherland model which interchanges
the two models with coupling constants  $\la$ and $1/\la$
\ref\Ga{M.Gaudin, Saclay preprint SPhT92-158}\ \ref\St{R.P.Stanley,
Adv.Math. 77 (1989) 76}.

In section two we recall how two complex fermions separate in two
independent semions, via the bosonisation.
In section three we show that the states of the $\widehat
{su(2)}_{k=1}$ theory can be represented in terms of
spinons. In section four we prove that the operators defined in
\nous\ form a representation of the Yangian algebra. In section four we
identify the irreducible sub-representations of this algebra and in section
six we construct the highest weight vectors of these representations.
Section seven contains the expressions of the first three conserved charges
commuting with the Yangian generators.

\newsec{Spinon-Holon separation in CFT.}

\def\upa{ \uparrow}
\def\dow{\downarrow}
Consider two complex fermions $\Psi_\sig$ and $\Psi_\sig^{\dag}$,
$\sig=\upa,\ \dow$, with operator product expansion~:
\eqn\EAa{
\Psi_\sig^{\dag}(z)\Psi_{\sig'}(w) \sim
\frac{\de_{\sig;\sig'}}{z-w} }
After Bardaki and Halpern \ref\Bar{K.Bardakci and M.Halpern,
Phys.Rev.D 3 (1971) 2493},
 it is well known that the currents
$J^a(z)$, bilinear in the fermions and defined by,
\eqn\EAb{
\cases{ J^+ = :\Psi_\upa^{\dag} \Psi_\dow: &~\cr
J^- = :\Psi_\dow^{\dag} \Psi_\upa: &~\cr},\quad
J^0 = :\Psi_\upa^{\dag} \Psi_\upa: - : \Psi_\dow^{\dag} \Psi_\dow:}
form a representation of the current algebra $\hat {su(2)}$
at level $k=1$. This representation is highly reducible in the
fermionic Fock space. In particular, the currents $J^a(z)$ commute
with the fermion number current. What is less familiar, but
which is a direct generalisation of the $so(4)$ symmetry of the
Hubbard model \ref\Ya{C.N.Yang and S.C.Zhang, Mod.Phys.Lett.B
4 (1990) 759}, is that there exists another set
of $\hat {su(2)}$ currents $K^a(z)$.
They are defined by,
\eqn\EAc{
\cases{K^+ = :\Psi_\upa^{\dag} \Psi_\dow^{\dag}: &~\cr
K^- = :\Psi_\upa \Psi_\dow: &~\cr},\quad
K^0 = :\Psi_\upa^{\dag} \Psi_\upa: + : \Psi_\dow^{\dag} \Psi_\dow:}
Moreover, these currents commute with the currents \EAb.
The simplest way to realize it consists in bosonizing the fermions.
Let us introduce two real bosonic fields $\varphi_\sig(z)$, $\sig=\upa,\ \dow$,
normalized to,
\eqn\EAd{
\vev{\varphi_\sig(z)\varphi_{\sig'}(w)} = \de_{\sig,\sig'}
\log(z-w) }
The bosonization rules are~:
\eqn\EAe{
i\d_z\varphi_\sig = :\Psi_\sig^{\dag} \Psi_\sig:\ ,\quad
\cases{\Psi_\sig^{\dag} = e^{i\varphi_\sig} &~\cr
\Psi_\sig = e^{-i\varphi_\sig} &~\cr}}
Define now the ``spin" and ``charge" bosonic sectors by
introducing the normalized fields $\varphi_s$ and $\varphi_c$,
\eqn\EAf{
\varphi_s = \inv{\sqrt{2}}\({ \varphi_\upa -\varphi_\dow}\),\quad
\varphi_c = \inv{\sqrt{2}}\({ \varphi_\upa +\varphi_\dow}\) }
The currents \EAb~ and \EAc~ are then given by~:
\eqn\EAg{
\cases{ J^\pm = e^{\pm i \sqrt{2}\varphi_s}  &~\cr
	J^0 = i\sqrt{2}\d_z\varphi_s &~\cr }\ ,\quad
\cases{ K^\pm = e^{\pm i \sqrt{2}\varphi_c}  &~\cr
	K^0 = i\sqrt{2}\d_z\varphi_c &~\cr } }
This is the standard free field construction \ref\FK{I.Frenkel
and V.G.Kac, Invent.Math. 62 (1980) 23}\ \ref\Se{G.Segal,
Commun.Math.Phys. 80 (1981) 301}\
of the level one representation of $su(2)$.

Since by construction the spin and charge bosons commute,
the spin currents $J^a(z)$ and the charge currents $K^a(z)$
are also mutually commuting. Naively, the symmetry algebra would
thus be $\hat {su(2)} \times \hat {su(2)}$. But as we will
see below, there is $Z_2$ condition, similar to
what happens in the Hubbard model, which reduces the symmetry
to $\hat {so(4)} = \hat {su(2)} \times \hat {su(2)} / Z_2$.

After bosonization the fermions fractionalize in two commuting parts,
the spinon and the holon~:
\eqn\EAgg{
\cases{\Psi_{\upa}(z)=\phi_s^+(z)\phi_c^+(z) &~\cr &~\cr
       \Psi_{\dow}(z)=\phi_s^-(z)\phi_c^+(z) &~\cr}, \quad
\cases{\Psi_{\upa}^{\dag}(z)=\phi_s^-(z)\phi_c^-(z) &~\cr
       \Psi_{\dow}^{\dag}(z)=\phi_s^+(z)\phi_c^-(z) &~\cr}}
with $ \phi^{\pm}_{s,c}=e^{\pm i \inv{\sqrt{2}} \varphi_{s,c} }$
In sect. 6 we will show how the operators $\phi_s^\pm(z)$ create
the one spinon states. Eq. \EAgg\ expresses the fact that the
fermions fractionalize into particles satisfying a half-integer
statistics~: spinons and holons are ''semions" \Hal\
\ref\Hald{F.D.M.Haldane, Phys.Rev.Lett. 67 (1991) 937}.

The fermionic Fock space is now finitely reducible for the
$\widehat{so(4)}$ affine algebra. For concretness, let us consider
the case for which both fermions are in the Neveu-Schwarz
sector. The partition function is then~:
\eqn\EAh{
Z_{{\rm NS}^2} = {\rm Tr}\({q^{L_0}}\)
=\prod_{n>1}\({1+q^{n-\half}}\)^4
= \inv{\eta(q)^2} \sum_{p_\upa,p_\dow \in Z}
q^{\half(p_\upa^2+p_\dow^2)} }
Here, $\eta(q)=\prod_{n\geq 0}\({1-q^n}\)$.
In the second equation we used the bosonization formula \EAe.
Introduce now the spin and charge zero modes,
$p_s=\inv{\sqrt{2}}(p_\dow-p_\upa)$ and
$p_c=\inv{\sqrt{2}}(p_\dow+p_\upa)$. Since in the ${\rm NS}^2$
sector the zero modes $p_\sig$ are both integers, the
momentum $(p_s\sqrt{2})$ and $(p_c\sqrt{2})$ are always both even or both odd.
Taking this $Z_2$ constraint into account, we find~:
\eqn\EAk{
Z_{{\rm NS}^2} = \La^{(s)}_0(q) \La^{(c)}_0(q)+
\La^{(s)}_1(q) \La^{(c)}_1(q)}
where $\La_j$, $j=0,\ 1$ are the character of the level one
representation of $\hat {su(2)}$~:
\eqn\EAl{
\La_j(q) = \inv{\eta(q)} \sum_{p\in Z +j/2} q^{p^2} }

For the Haldane-Shastry spin chain, the number of fermions
per site is exactly one, so the charge degrees of freedom
are frozen. In the conformal limit, this constraint is
equivalent to the condition~:
$$ K_n \ket{phys} =0,\quad \forall n\ge0.$$
which leads us to consider only one of the two $\widehat{su(2)}$
subalgebras.

\newsec{Spinon representation of $\hat {su(2)}$.}

The low-energy limit of the Haldane-Shastry spin chain gives
a description of the level one representation of $\hat {su(2)}$
in terms of spinons \nous.
To each state is associated a motif
which is a semi-infinite sequence of symbols $0$ and $1$ with
the rules that a) two $1$ can never be adjacent
b) the sequence stabilizes at infinity to a sequence
of alternating $0$ and $1$.
and c) the first figure of the sequence is a $0$.
There are two possible asymptotic behaviors: the symbols
$1$ are either on even or on odd positions.
These two asymptotics correspond to the two level one
representations of $\hat {su(2)}$. The sequence
$01010101\cdots$ corresponds to the vacuum, and the
sequence $001010101\cdots$ to the spin half primary field.
The excited states are therefore described by finite
rearrangement of these two primary sequences.
A motif can be coded into the sequence of
integers $\{m_j\}$ indicating the positions of the symbols $1$.
For example the vacuum sequence is $\{m_j^{(0)}=2j+1,\ j\geq 0\}$,
and the spin one half primary sequence is $\{m_j^{(1)}=2j+2,\ j\geq 0\}$.

A given motif corresponds to a multiplet of states. This degeneracy
is best described in terms of spinons. We refer to
the interval between the $1$'s in position $m_k$ and $m_{k+1}$
as the spinon orbital of momemtum $k$.
$(Q+1)$ consecutive $0$ in the $k^{\rm th}$ orbital represent $Q$
spinons of momentum $k$, or equivalently,
the total occupation number of the $k^{\rm th}$
orbital is $n_k=m_{k+1}-m_k -2$.
The degeneracy of a motif is recovered by considering the spinons
in each orbital as spin $\half$ bosons.
Therefore, in each orbital there is an independent $su(2)$
symmetry under which the spinons are spin $\half$ objects.

Let us give an alternative description of the degeneracy of a
multiplet which has the advantage of being generalisable to
the finite chain as well as to the $\hat {su(n)}$ case.
Consider the configurations of a semi infinite chain of spins
taking the values $+$ or $-$. To a configuration we assign a
sequence of $0$ and $1$ by applying the following rules:
Put a $0$ between consecutive spins if the second spin is
less or equal than the first one, and put a $1$ if the second
spin is larger than the first one.
Also put a $0$ in front of the first spin of
the chain. Following these rules, two $1$ cannot be consecutive.
The number of spin configurations which correspond to a motif of $0$ and $1$
gives the number of states in the multiplet of this motif.
This description of the degeneracies is very similar
to the cristal basis used in quantum groups
\ref\Ka{M.Kashiwara, Commun.Math.Phys. 133 (1990) 249}.
To have a better understanding of this connection, one would need
the q deformation of the spin chain hamiltonian \YB .

The Virasoro generator $L_0$ acts on the motifs by
$L_0=\sum_i\( m_i - m_i^{(j)}\)$ where $m_i^{(j)}$, $j=0,1$
is the appropriate primary sequence.
In the spinon description it becomes~:
\eqn\EBa{
L_0= \({\frac{N_{sp}}{2}}\)^2 + \sum_{i=1}^{N_{sp}} k_i
= \({\frac{N_{sp}}{2}}\)^2 + \sum_{k\geq 0} k\ n_k }
where $n_k$ is the occupation number of the $k^{\rm th}$ orbital,
and $N_{sp}$ the total occupation number: $N_{sp}=\sum_k n_k$.
In the vacuum representation the number of spinons $N_{sp}$
is even, while in the spin half representation it is always odd.
In other words, in a given sector the spinons can only be
created by pairs.

Let us now compute the characters using the spinon description.
Using \EBa~ and the fact that the spinons are spin $\half$ objects,
we have~:
\eqn\EBb{
\La(q;z) = {\rm Tr}\({ q^{L_0}\ z^{J^0_0} }\)
= \sum_{\{ n_k\}} z^{\frac{(n^+-n^-)}{2}}
q^{(n^++n^-)^2/4} q^{\sum_kk(n_k^++n_k^-)} }
where $n_k^\pm$ are the up and down spinon occupation numbers,
and $n^\pm = \sum_k n_k^\pm$.
Using the identity,
\eqn\EBc{
\sum_{\partkN} q^{\sum_{i=1}^N k_i}
= \prod_{l=1}^N \sum_{m_l=0}^\infty q^{lm_l} }
we can write eq. \EBb~ as~:
\eqn\EBd{ \La(q;z)
= \sum_{n^+,n^-} z^{\frac{(n^+-n^-)}{2}} \
\frac{q^{(n^++n^-)^2/4}}{(q)_{n^+}(q)_{n^-}} }
with $(q)_n= \prod_{j=1}^n(1-q^j)$. This quasi-particle
formula for the characters
recently appeared in \ref\Me{E.Melzer, preprint TAUP 2125-93
hep-th/9312043}.

\newsec{Yangian invariance of $\hat {su(2)}$.}

The Haldane-Shastry $su(2)$ spin chain has an infinite symmetry, the
$su(2)$ Yangian. This suggests that the conformal field theory which
describes the low energy limit of this model possesses the same symmetry.
The first two generators of the Yangian, which are sufficient to
determine all the others, were defined in \nous ~:
\eqn\ECa{\eqalign{
&Q_0^a = J_0^a\cr \noalign{\vskip3pt}
Q_1^a =g \oint\limits_{|z_1|>|z_2|} \frac{dz_1}{\2pi} \oint\limits_{C_0}
&\frac{dz_2}{\2pi} \ \th_{12} \ f^{abc}
: J^b(z_1)J^c(z_2):\cr}}
where $\th_{12} = {z_1}/ {(z_1-z_2)} $, $C_0$ is a contour around the
origin and $g$ a normalization constant
to be fixed later. The double dots represent the normal ordering
operation. The field $J^a(z)$ are defined
by their Fourier expansion $J^a(z)=\sum_n J_n^a z^{-n-1}$ with the
modes $J_n^a$ being the generators of the Kac-Moody algebra:
$\[J_n^a,J_m^b\]=f^{abc}J_{n+m}^c+kn\de_{n+m,0}\de^{ab}$. In terms of
these modes~:
\eqn\ECb{
Q_1^a =\ g f^{abc} \sum_{m>0} :J_{-m}^b J_m^c:}
We define the action of these generators on a product of primary fields
by the following contour integrals~:
\eqn\ECc{\eqalign{
Q_0^a\( \prod_{i=1}^N \phi_i\(w_i\)\)& = \oint\limits_{C_i}
\frac{dz}{\2pi}\ J^a(z)\(  \prod_{i=1}^N \phi_i(w_i)\)~\cr
Q_1^a\( \prod_{i=1}^N \phi_i(w_i)\) =\ g \oint\limits_{|z_1|>|z_2|}
\frac{dz_1}{\2pi}& \oint\limits_{C_i} \frac{dz_2}{\2pi}\
\th_{12}\ f^{abc}
:J^b(z_1)J^c(z_2):\( \prod_{i=1}^N \phi_i(w_i)\)~ \cr}}
with the contours of integration $C_i$ encircling the points $w_i$.
The fields $\phi$ are $su(2)$ vectors~: $\phi(w)=\phi^+(w)e_++
\phi^-(w)e_-$, with $(e_+,e_-)$ a basis of $C^2$.

In order to evaluate the integrals in \ECc\ we need the operator product
expansion:
\eqn\ECd{
J^a(z)\phi(w) = \frac{J_0^a \phi(w)}{z-w}+J_{-1}^a
\phi(w)+\ldots}
For the $\hat {su(2)}\ k=1$ case, there is only one independent
descendant field at the first level. This permits us to write
the second term in \ECd\ as:
\eqn\ECe{
J_{-1}^a \phi(w) = \frac{1}{2\De}\ t^aL_{-1}\phi(w) =
\frac{1}{2\De}\ t^a \d_w \phi(w)}
where $t^a$ are the transposed of the $su(2)$ matrices in the fundamental
representation, $\[t^a,t^b\]=-f^{abc}t^b$,
and $\De=1/4$ is the conformal dimension of the primary fields $\phi$.

It is more convenient to evaluate \ECc\ on the components $\phi^{\al}$
of the fields $\phi$
$$Q_n^a\ \pphia \equiv \bra{e^{\al_1},\ldots,e^{\al_N}} Q_n^a \ket{
\prod_{i=1}^N\phi(w_i)}$$
To fix our normalization, we take $f^{abc}=2i\epsilon^{abc}$. It
follows that the quadratic Casimir in the adjoint representation
$(f^{abc}f^{bcd}=c_v \de^{ad})$ has the value $c_v=-8$.
If $\sig^a \sig^a=c_g I$, then $c_g=3$. For later convenience, we take
$g=4\De/c_v=-1/8$.

After performing the integrals in \ECc\ we obtain the matrix elements
of the operators $Q^a$ :
\eqn\ECf{\eqalign{
&Q_0^a\ \pphia= \(\sum_{j=1}^N\ \sig_j^a \) \pphia \cr
Q_1^a\ \pphia = & \[ \sum_{j=1}^N \(\inv{2} + w_j \d_{w_j}\) \sig_j^a
-\inv{c_v}\sum_{j\not=k}\th_{kj}\ f^{abc} \sig_k^b \sig_j^c \]
\pphia \cr}}
The operators $\sig_j^a$ act on the fields $\phi_j$ as follows~:
\eqn\ECff{\eqalign{
&\sig_j^0 \phi_j^{\pm}(w_j) =\pm \phi_j^{\pm}(w_j) \cr
&\sig_j^+ \phi_j^+(w_j) =\sig_j^- \phi_j^-(w_j)=0 \cr
&\sig_j^+ \phi_j^-(w_j) = \phi_j^+(w_j) \cr
&\sig_j^- \phi_j^+(w_j) = \phi_j^-(w_j) \cr }}
with $\sig_j^{\pm}=1/2(\sig_j^1\pm\sig_j^2)$. These operators have
the commutation relations
$\[\sig_i^a,\sig_j^b\]=\de_{ij}f^{abc}\sig_i^c$.

In \ECf\ one recognizes the Yangian generators for the $su(2)$
dynamical model. As defined in \YB , the coupling
constant has the value $ \la = -1/2 = - 2 \Delta $. This implies that
the operators \ECa\ form a representation of the $su(2)$ Yangian.
Note that the fact that the Yangian representation \ECf\ is invariant by
simultaneous permutations of the positions and spin $(w_i, \sig_i^a)$
is a direct consequence of the contour integral definition \ECc .
For an algebra other than $\hat {su(2)}$, $k=1$, the equation \ECe\
does not hold and the definition \ECa\ of the Yangian
generators must be changed \ref\Sch{K.Schoutens, preprint
PUPT-1442 hep-th/9401154}.

A consequence of the Yangian symmetry is the existence of
Ward identities~:
\eqn\ECg{\eqalign{
\sum_{i=1}^N \vev{\ \sig_j^a \pphia\ }&=0 \cr
\sum_{i=1}^N w_j\d_{w_j} \vev{\ \sig_j^a \pphia\ }-\inv{c_v}\
\sum_{j\not=k}\th_{kj}&\ f^{abc} \vev{\ \sig_k^b\sig_j^c \pphia\ }=0 \cr}}
The two point correlation functions are solutions of these
equations.

Starting with the first two generators, $Q_0^a$ and $Q_1^a$,
we can construct a transfer
matrix satisfying  the Yang-Baxter equation \ref\Fa{L.D.Faddeev,
{\it''Integrable models in 1+1 dimensional quantum field theory"}
Les Houches lectures, Elsevier Science Publishers (1984)}~:
\eqn\ECi{
R_{00'}(x-y)\ T^0(x)\ T^{0'}(y) = T^{0'}(y)\ T^0(x)\ R_{00'}(x-y)}
with $R(x-y) = (x-y)+ \la P_{00'}$, $P_{00'}$ being the permutation operator
which
exchange the two auxiliary spaces $0$ and $0'$. If we denote by
$T(x)$ a matrix with operator entries, $T^{ab}(x)$,
then $T^0(x)=\sum_{a,b} X^{ba}_0 T^{ab}(x)$ with $X^{ba}_0$
the matrices acting as $\ket{b} \bra{a}$ in the auxiliary space $0$.
The generators of the Yangian algebra are the coefficients in the
asymptotic expansion of $T(x)$~:
\eqn\ECj{
T^{ab}(x) = \de^{ab}+\la \sum_{n\ge0} \inv{x^{n+1}} T_n^{ab} }
The equation \ECi\ is equivalent to the following commutation
relations~:
\eqna\ECn
$$\eqalignno{
[T^{ab}_{n+1},T_m^{cd} ] -[T_n^{ab} ,&T_{m+1}^{cd} ]
=\la \( T_n^{cb}T_m^{ad}-T_m^{cb}T_n^{ad}\) &\ECn a\cr
[T_0^{ab},T_n^{cd}] &=\de^{cb}T_n^{ad}-\de^{ad}T_n^{cd} & \ECn b\cr}$$
The transfer matrix is uniquely determined by giving the first two
generators and imposing the condition \ref\Dr{V.G.Drinfeld,
{\it''Quantum Groups"}, Proc. of the ICM, Berkeley (1987) 798}\
\ref\Dri{V.G.Drinfeld, Funct.Anal.Appl. 20 (1988) 56}~:
\eqn\ECk{
Det_qT(x) = T^{11}(x)T^{22}(x+\la)-T^{12}(x)T^{21}(x+\la) = 1}
The quantum determinant $Det_qT(x)$ commute with all the
elements in the algebra \ref\Kor{A.G.Izergin and V.Korepin,
Sov.Phys.Dokl. 26 (1981) 653}, so the condition \ECk\ can be
imposed in a consistent way.
The traces of $T_n$ are not restricted by \ECn\ ; to fix them
one have to impose \ECk\ order by order in $x$.
For the first two orders, the result is~:
\eqn\ECm{\eqalign{
&T_0^{11}+T_0^{22} = 0 \cr
T_1^{11}+T_1^{22} &=\ \la\(T_0^{12}T_0^{21}-T_0^{11}T_0^{22}+T_0^{22}\) \cr}}

Our definition for the matrix elements of $T_0$ and $T_1$ is~:
\eqn\ECl{
T^{12}_n = Q_n^+,\quad T_n^{21} = Q_n^-,\quad T_n^{11}-T_n^{22} = Q_n^0
,\qquad n = 0,1}
We take \ECn a , $n>0,\ m=0$ as the definition of the higher level
generators. The traces $\sum_aT_n^{aa}$ are fixed using the
procedure we have just mentioned. The value of the coupling
constant consistent with this definition is $\la=-1/2$.

A redefinition of the generators ${Q'}_0^a=Q_0^a\ ,\ {Q'}_1^a=Q_1^a
+\al Q_0^a$ is always possible. The only effect on the transfer matrix
is a spectral parameter shift $T'(x)=T(x-\al)$.
The construction of the transfer matrix with  unit determinant
from the first two generators is unique and the action of
our input generators \ECl\ is the same as for the
dynamical model transfer matrix generators. These observations
enable us to write~:
\eqn\ECo{
T_{cft}(x)\pphia=f(x)\ T_{dyn}(x-\frac{N+1}{4})\pphia }
Here $T_{cft}$ is the matrix we have just defined,  $T_{dyn}$
is the transfer matrix for the N-particle dynamical model at
$\la=-1/2$ and $f(x)$ is a function with the property
$Det_q\(f(x)T_{dyn}(x)\)=1$.

The dynamical transfer matrix, for the coupling constant $\la=-1/2$,
was defined in \YB ~:
\eqn\ECp{
T_{dyn}^{ab}(x)=\de^{ab}-\inv{2}\sum_{n\ge0}\sum_{j,k=1}^N
X_j^{ab}(L^n)_{jk}\ x^{-n-1} }
with $L_{ij}=w_j \d_{w_j} \de_{ij}-\inv{2}(1-\de{ij})\th_{ij}X_i^{ab}
X_j^{ba}$ and $X_i^{ab}$ acting as $\ket{a}\bra{b}$ on
$i^{\rm th}$ $su(2)$ copy. Here, the action of $X_i^{ab}$ on a primary field
is, according to the definition \ECff , $X_i^{ab}\phi_i^c=\de^{bc} \phi_i^a$.
On functions symmetric by simultaneous permutations of coordinates
and $su(2)$ indices, say $\psi(w,\al)$, the action of
the transposed transfer matrix $T^0_{dyn}(x)=X_0^{ab}T^{ba}_{dyn}$
is given by~:
\eqn\ECq{
T^0_{dyn}\psi(w,\al)=\prod_{j=1}^N\(1-\inv{2}\frac{X_0^{ab}X_j^{ba}}
{x-\Dj} \)\psi(w,\al)}
where 0 is the auxiliary $su(2)$ space.
The differentials $\Dj$ are defined
\ref\Du{C.F.Dunkl, Transl.Amer.Math.Soc. 311 (1989) 167}\
\ref\Op{E.M.Opdam, Invent.Math. 98 (1989) 1}\
\ref\He{G.E.Heckman, Invent.Math. 98 (1991) 341}\
\Che\ \Po\ by~:
\eqn\ECr{
\Dj=w_j\d_{w_j}-\half{\sum_{i>j}\th_{ji}K_{ji}}+\half{\sum_{i<j}
\th_{ij}K_{ji}} }
where the operators $K_{ij}$ permute the coordinates, $K_{ij}w_i=w_jK_{ij}$.
Note that \ECq\ follows from \ECp\ if the permutation operators
$P_{ij}=\sum_{ab} X_i^{ab}X_j^{ba}$ are replaced with the position
permutations $K_{ij}$.

The operators $\(\sN\hat D_i^n\)$ commute with the dynamical transfer
matrix \ECq\ and they are diagonal on the following
basis of symmetric functions~:
\eqn\ECs{
\psi_{\partk} \vecw=\pref\  P_{\partk}^{(-1/2)}\vecw}
Here, $P_{\partk}^{-1/2}\vecw$ denotes the Jack (Sutherland) polynomial,
symmetric in $w$, associated to the partition $\partk$, with $\ k_1\ge
k_2 \ge \ldots \ge k_N$, at the value of the parameter $\la=-1/2$
\ref\Su{B.Sutherland, Phys.Rev.A 5 (1972) 1372}\ \St ,
Starting from the integers $k_i$, we define a set
of half-integers, $d_i$~:
\eqn\ECt{
d_i=k_i+\inv{2}(N-i)}
which characterize the eigenvalues of $\sN\hat D_i^n$~:
\eqn\ECu{
\sN\hat D_i^n \ \psi_{\partk} \vecw=\(\sN d_i^n\)\ \psi_{\partk}\vecw}

\newsec{The spinon multiplets.}

We now want to identify the irreducible sub-representations of the
Yangian algebra in the level one representation of $\widehat{su(2)}$.
The first step is to find the highest weight
vectors, i.e. the vectors satisfying~:
\eqn\ECv{\eqalign{
T^{12}(x) \ket{h.w.} &=0 \cr
T^{11}(x) \ket{h.w.} &=t^{11}(x) \ket{h.w.} \cr}}
where $t^{11}(x)$ is a function and not an operator. The other
vectors in the representation are obtained by repeated
actions of the lowering operator $T^{21}$.
Obviously, the product of fields $ \prod_{i=1}^N \phi_i^+\(w_i\)$
is annihilated by $T^+(x)=T^{12}(x)$, so this product is a
generating function for the highest weight vectors.
We shall represent the primary fields $\phi$ as vertex operators~:
\eqn\ECx{
\phi^{\pm}(z)=:e^{\pm i \inv{\sqrt 2}\vphi(z)}: }
where $\vphi(z)$ is the bosonic field
$\vphi(z)= q-ip\ln z +i\sum_{n\not=0} \al_n \frac{z^{-n}}{n}$.
The product of vertex operators has the property~:
\eqn\ECy{
\pphip = \pref :e^{ i \inv{\sqrt 2}\sN \vphi(w_i)}: }
When acting on the vacuum state, only the terms regular at
$w_i=0$ in \ECy\ will survive. As the normal ordered exponential
is monovalued and symmetric in $w_i$, we can expand this product on
a basis of symmetric function. We choose the basis of the polynomials
defined in \ECs ~:
\eqn\ECz{
\pphip \rvac =\pref \sum_{l(k)=N} P_{\partk}^{(-1/2)} \vecw \ket{\partk} }
where $l(k)$ is the length (number of parts) of the
partition $\partk$.
Using the representation \ECq\ of the transfer matrix, the property
\ECu\ of the Jack polynomials and the fact
that the basis in \ECz\ is orthogonal, we infer the action
of $T_{cft}(x)$ on the states $\ket{\partk}$~:
\eqn\ECw{\eqalign{
T_{cft}(x)\ket{\partk} &=f(x)
\ \pmatrix{\prod\limits_{i=1}^N \(1-\inv{2} {1\over{x-
{{N+1}\over{4}}-d_i}}\)
&0\cr  \* & 1 \cr} \ket{\partk} \cr
&=f(x)\ \pmatrix{\frac{P\(x-\inv{2}\)}{P(x)} & 0 \cr
\* & 1 \cr} \ket{\partk} \cr }}
Thus, for each highest weight vector $\ket{\partk}$ there is an
associated polynomial $P(x)=\prod_{i=1}^N \(x-\frac{N+1}{4}
-d_i\)$. According to Drinfeld \Dr\ \Dri ,
all the information about
the corresponding representation is encoded by the
positions of the roots $d_i$ of this polynomial.

In order to obtain the $su(2)$ content of the representation
labelled by $\partk$ (or $\lbrace d_i \rbrace$) we use a
canonical transfer matrix~:
\eqn\ECza{
T_d^{ab}(x)=\de^{ab}-\inv{2} \frac{X^{ab}}{x-d} }
that satisfy the $RTT=TTR$ equation, with $X^{ab}$ the $su(2)$
matrices in the fundamental representation. Take the coproduct
of two such matrices~:
\eqn\ECzb{
T^{ab}_{d_1,d_2}(x)=\sum_c T_{d_1}^{ac}(x) \oti T_{d_2}^{cb}(x)}
The h.w. vector of this matrix is the tensor product of the
h.w. vectors of elementary matrices $T_d$, $\ket {++}=\ket{+}
\oti \ket{+}$, and has the spin equal to 1. If $d_1=d_2+1/2$,
the raising and lowering operators $T^{\pm}_{d_1,d_2}(x)$ will
always send a spin 1 (symmetric) vector on a spin 1 vector. As
the spin 0 state is not attainable from the h.w. state (and is not
a h.w. state), we can drop it out.
If $d_1\not=d_2+1/2$, the matrix \ECzb\ corresponds to a product of two
spin $1/2$ representations.
In the same way, consider the matrix~:
\eqn\ECzc{
T_{d_1,\ldots,d_N}(x)= \bigotimes\limits_{i=1}^N T_{d_i}\(x-\frac{N+1}{4}\)}
Any sequence of $Q$ $d_i$'s separated by $1/2$ corresponds to a spin
$Q/2$ $su(2)$ representation, and the matrix \ECzc\ corresponds
to the $su(2)$ representation resulting from the tensor product
of these spin $Q/2$ representations. Moreover,
\eqn\ECzd{
T_{d_1,\ldots,d_N}(x) \ket{h.w.} =
\pmatrix{ \frac{P\(x-\inv{2}\)}{P(x)} & 0 \cr
	  \* & 1 \cr} \ket{h.w.} }
with $P(x)$ the same as in \ECw . In conclusion, the representation
generated from $\ket{\partk}$ and the representation induced
by the transfer matrix \ECzc\ are isomorphic.

In the following we show how to associate a motif to the highest
weight vector labelled by $d_i$'s $(d_i=k_i+\inv{2}(N-i)$).
Start with a set of empty cases labelled by natural numbers (including
zero). Put a $0$ on the positions $2d_1$,...,$2d_N$, then fill the other
cases with alternating $0$ and $1$. This is possible, since $2d_i-2d_{i+1}$
is always an odd number ($d_i=k_i+\inv{2}(N-i)$)
and the least of these numbers, $2d_N$,
is even. The resulting motif never has two $1$'s on adjacent positions.
The asymptotic behaviour of the motif is dictated by the parity
of $2d_1$, i.e. by the parity of $N$. If $N$ is even, the
motif is a finite rearrangement of the vacuum motif $01010101 \cdots$,
if $N$ is odd, the motif is generated from the primary
sequence $001010101 \cdots$.

The same construction can be achieved starting from the set $\partki$
. Take the semi-infinite sequence $01010101 \cdots$ and call orbital
the interval between two $1$'s separated by a sequence of $0$'s;
then, for each $k_i$,
$i=1,\ldots,N$ insert a $0$ in the $k_i^{\rm th}$ orbital.
This procedure gives the result described in sect. 3 (the $k_i$'s
are the same here and in sect. 3).

\newsec{Spinon creating operators.}

The one spinon states are generated by the positive
modes of the primary fields~:
\eqn\EDa{
\phi^{\pm}(w) \rvac = \sum_{n \ge 0} w^n\ \modn}
where $\modn$ denotes the states with the spinon momentum equal to $n$.
These states are characterized by the values of the charge and
energy~:
\eqn\EDb{
p\ \modn =\pm \inv{\sqrt 2}\ \modn, \qquad L_0\ \modn= (n+\inv{4})\ \modn}
Remark that for each $n$ there is a doublet of states ("spin-up"
and "spin-down" spinons). The corresponding motifs are of type~:
$01010 \cdots 10100101 \cdots$ . The lowest energy doublet has $n=0$ and is
represented by the motif $001010101 \cdots$.

Similarly, the $N$-spinon highest weight vectors are obtained from a
product of $N$ fields~:
\eqn\EDc{
\phi^+(w_1) \ldots \phi^+(w_N) \rvac = \pref \sum_{\partkN}
P_{\veck}^{(-1/2)} \vecw \ket{ \veck} }
where $P_{\veck}^{(-1/2)}$ is again the Jack polynomial associated
to the partition $\partkN$ at $\la=-1/2$.
The charge and energy of the $N$ spinon states are~:
\eqn\EDcc{\eqalign{
p\ \ket{\veck}&=\frac{N}{\sqrt2}\ \ket{\veck},\cr
L_0\ \ket{\veck}&=\[ \(\frac{N}{2}\)^2 + \sN k_i \] \ \ket{\veck}\cr }}

Note that the operators creating the $N$-spinon states $\ket{\partk}$
are not the product of $N$ operators creating the one-spinon states
$\ket{k_1},\ \ket{k_2},\ldots$. This fact is an echo of the semionic
statistics of the spinons.

One can try to obtain the states $\ket{\veck}$ from the vacuum
using the bosonic creation operators, $\al_n$. A method for doing this
is inspired from the construction of Jack polynomials. We need
to recall few basic properties of these polynomials. We refer to
\St\ for more details.
The Jack Polynomials $P_{\partk}^{(\la)} \vecw$
are the eigenfunctions of the differential operator~:
\eqn\EDj{
D(\la)=\sN w_i^2 {\d_{w_i}}^2-2\la \sum_{i\not=j} \frac{w_i^2}
{w_i-w_j} \d_{w_i} }
which is a gauge transformed version of the Calogero-Sutherland
hamiltonian.
They are indexed by a partition $\partk$ of length $l(k)\le N$
and the corresponding eigenvalues are~:
\eqn\EDk{
e_{\partk}=\sum_ik_i^2+\la \sum_j {k'}_j^2-[\la(2N-1)-1]\sum_i k_i }
with $k'_j$ the parts of the partition conjugate to $\partk$.
If we describe a partition $\partk$ as a Young tableau
with $N$ rows of $k_1,\ldots,k_N$
squares, then the conjugate partition $\partkp$ has $N$ columns of
$k_1,\ldots,k_N$ squares.
Let $w=\vecw$ be a set of variables
($N$ arbitrary), $\partk$ a partition of length $l(k)$ and
$|k|=\sum_ik_i$. Define $p_{\partk}$ as follows~:
$$ p_{\partk} = \prod_{i=1}^{l(k)} \ p_{k_i}
,\qquad p_n=\sum_{j=1}^N w_j^n$$
If $\partk$ has $n_i$ parts equal to $i$, then write~:
$$z_{\partk}=(1^{n_1}2^{n_2}\ldots)n_1!n_2!\ldots$$
Define a scalar product on the space of symmetric functions by~:
\eqn\EDdd{
\vev{\ p_{\partk},p_{\partr}}=\de_{\partk,\partr}\ z_{\partk}
\ (-\la)^{-l(k)} }
where $\de_{\partk,\partr}$ is equal to $1$ if $\partk=\partr$ and is
zero otherwise. The Jack polynomials also are orthogonal with
respect to this scalar product.
The power-sum symmetric functions $p_{\partk}$ and Jack symmetric
functions $P_{\partk}^{(\la)} $
provide two basis for the space of symmetric functions, connected by
a triangular matrix~:
\eqn\EDe{
P_{\partk}^{(\la)} =\sum_{\partr \le \partk} c_{\partk \partr}
(\la)\  p_{\partr} }
where the partial ordering on the partitions is defined as follows~:
$\partr \le \partk $ if $|r|=|k|$ and $r_1+r_2+\ldots+r_i \le
k_1+k_2+\ldots+k_i$ for any $i$.
The constants $c_{\partk \partr}$ are calculable in principle,
but there is no general formula for them. We give as examples the
expressions of $P_{\partk}^{(\la)}$ for $ |k|=1,2,3 $ normalized
such that the coefficient of the monomial $z_1^{k_1}z_2^{k_2}\ldots$
is equal to $1$~:
\eqn\EDee{
P^{(\la)}_1=p_1,  \quad
\cases{P^{(\la)}_{1,1}=\half(p_1^2-p_2) \cr
       P^{(\la)}_2=\frac{\la}{\la-1}\(p_1^2-\inv{\la}p_2\) \cr} ,\quad
 \cases{P^{(\la)}_{1,1,1}=\inv{6}(p_1^3-3p_1p_2+2p_3) \cr
  P^{(\la)}_{2,1}=\frac{\la}{2\la-1}
  \(p_1^3-\frac{1+\la}{\la} p_1p_2+\frac{1}{\la} p_3\) \cr
  P^{(\la)}_3=\frac{\la^2}{(\la-1)(\la-2)}
  \(p_1^3-\frac{3}{\la} p_1p_2+\frac{2}{{\la} ^2}p_3\) \cr}}
A very important property of the Jack polynomials is the
duality $\la \leftrightarrow 1/\la$. Let $\omega_\la$ be an
automorphism on the ring of symmetric polynomials, defined by~:
\eqn\EDD{
\omega_{\la}(p_n)=\frac{(-1)^n}{ \la}\ p_n  }
Then~:
\eqn\EDE{
\omega_{\la}\(P^{(\la)}_{\partk}\)=j_{\partk}(\la)\
P_{\partkp}^{(1/\la)} }
where $\partk$ and $\partkp$ are two conjugate partitions and
$j_{\partk}(\la)$ is the norm of $P^{(\la)}_{\partk}$ with
respect to the scalar product \EDdd .
Now, consider two sets of variables, $w=\vecw$ and $t=(t_1,\ldots,t_M)$.
The duality property allows us to write one of the generating functions
of the Jack polynomials $(\la=-2)$ as~:
\eqn\EDd{\eqalign{
\prod_{i=1}^N \prod_{a=1}^M (1-t_aw_i) & = e^{-\sum_{n\ge0} \sN
\sum_{a=1}^M \frac{(t_aw_i)^n}{n}}  \cr
& =e^{-\sum_{n\ge0} \sN \frac{p_nw_i^n}{n}} = \sum_{\partk}(-1)^{|k|}
\ P^{(-1/2)}_{\partk}(w)\ P^{(-2)}_{\partkp}(t) \cr }}
The sum is over the partitions with at most $N$ parts with $k_1 \le M$.
The relation \EDe\ is independent of the number of variables.

We now apply eq. \EDd\ to the construction of the spinon creating
operators.
If we put formally $M\rightarrow \infty$, $\al_{-n}=-\sqrt 2 p_n$
for $n>0$, \EDd\ is the expansion of the positive modes of the
vertex operator product. In virtue of equation \EDc\ and taking into
account the zero mode, $q$, we obtain~:
\eqn\EDf{
\ket{\partk} = P^{(-2)}_{\partkp}(\al)\ e^{i \frac{N}{\sqrt 2} q}
\rvac }
Note that the scalar product \EDdd\ for $\la=-2$ is consistent
with the scalar product on the Fock space~:
\eqn\EDg{
\vev{\prod_i\al_{-n_i},\prod_j\al_{-m_j}}=\lvac \prod_i\al_{n_i}
\prod_j\al_{-m_j} \rvac }

\newsec{Conserved charges}

For the dynamical model, the conserved charges commuting
with the Yangian are generated by the quantum determinant of the
transfer matrix.
We denote by $H_n^{dyn}$ the conserved quantities of the dynamical
model. On symmetric functions, these operators act as
$\(\sN\hat D_i^n\)$, so their eigenvalues are $\sN \(k_i+\half (N-i)\)^n$.
As we have already imposed restrictions on the
quantum determinant of the conformal field theory tranfer matrix,
we cannot use the same procedure to obtain the conserved quantities
commuting with the Yangian generators \ECa .
At present, there is no systematical way to find the higher
order hamiltonians of the conformal field theory. However, we know
some operators which act on the primary fields as linear combinations of
the dynamical model hamiltonians.

Given a motif, we define $E_n=\sum_i\(m_i^{(j)}\)^n-\sum_i\(m_i\)^n$,
$m_i$ being the positions of $1$'s on the motif
and $m_i^{(j)}$, $j=0,1$ the positions of $1$'s on the appropriate
primary motif (as defined in sect.3).
As in \nous , we will express the values of the conserved
quantities in terms of $E_n$'s.

Obviously, the energy $L_0$ is one of the conserved charges~:
\eqn\EEa{\eqalign{
L_0&=\inv{\kappa} \oint\limits_{C_0} \frac{dz}{2\pi} \ z :J^a(z)J^a(z): \cr
L_0 \pphi \rvac &= \sN \(\dw +\De\) \pphi \rvac \cr
&= \(H_1^{dyn}+\frac{N}{4}\) \pphi \rvac \cr }}
with $\kappa=c_g/\De=12$ in our normalisation and $C_0$ a contour
that encircles the origin.
The energy evaluated on a highest weight state $\ket{\partk}$
is~:
\eqn\EEb{
L_0 \ket{\partk}= \(\sN\(k_i+\half (N-i)\)+\frac{N}{4} \)
\ket{\partk}=\(\(\frac{N}{2}\)^2+ \sN k_i \) \ket{\partk} }
The values of $L_0$ are the same as the values of $E_1$.

The second hamiltonian was defined in \nous as~;
\eqn\EEc{
H_2^{cft}= \oint\limits_{C_0} \frac{dz_1}{\2pi} \oint\limits_{| z_1|
> | z_2 |} \frac{dz_2}{\2pi} \ \th_{12} \th_{21}: J^a(z_1)J^a(z_2): }
The action of $H_2^{cft}$ on a product of primary fields is~:
\eqn\EEd{\eqalign{
H_2^{cft} \pphia \rvac &=\[ 4 \sN (\dw)^2+2\sN \dw -\sum_{i\not=j}
h_{ij} \sig^a_i \sig^a_j \] \pphia \rvac \cr
&= \(4H_2^{dyn}+2H_1^{dyn} \) \pphia \rvac \cr}}
with $h_{ij}=\th_{ij}\th_{ji}$.
For the highest weight vector, we find~:
\eqn\EEe{
H_2^{cft} \ketk =2\(2 \sN k_i^2-\sum_{j=1}^{k_1} {k'}^2_j
+2N\sN k_i + \inv{6}N(N^2-1) \) \ketk }
where $k'_j$ are the parts of the partition conjugate to $\partk$. The
eigenvalues of $H^{cft}_2$ are equal to the values of $2E_2$.

Like the second hamiltonian, the third one was inspired from
one of the conserved charges of the Haldane-Shastry spin chain~:
\eqn\EEf{\eqalign{
H_3^{cft}&= \oint\limits_{|z_1|>|z_2|} \frac{dz_1}{\2pi} \oint\limits_{
|z_2|>|z_3|} \frac {dz_2} {\2pi} \oint\limits_{C_0} \frac{dz_3}{\2pi}
\ \th_{12} \th_{23} \th_{31} \ f^{abc}:J^a(z_1) J^b(z_2) J^c(z_3): \cr
\noalign{\vskip5pt}
&+ \frac{3c_v}{2} \oint\limits_{|z_1|>|z_2|} \frac{dz_1}{\2pi}
\oint\limits_{C_0}
\frac{dz_2}{\2pi}\ \th^2_{12} \th_{21}\ :J^a(z_1)J^a(z_2): \cr }}
On a product of primary fields, this operator acts as~:
\eqn\EEg{\eqalign{
H_3^{cft} \pphia & =
\BBL \sum_{i \not=j \not=k } \th_{ij} \th_{jk}
\th_{ki}  f^{abc} \sig_i^a \sig_j^b \sig_k^c
-12\sum_{i\not=j} h_{ij}(\dw+w_j\d_{w_j}) \sig_i^a \sig_j^a \cr
 +32\sN(\dw)^3 & +48\sN (\dw)^2 +16\sN \dw
-12 \sum_{i\not=j} h_{ij} \sig_i^a \sig_j^a
\BBR \pphia \cr
& =\(32H_3+48H_2+16H_1\)^{dyn} \pphia \cr}}
The eigenvalues of this operator can be calculated from the eigenvalues
of the operators $H_n^{dyn}$. The same values are obtained if we
evaluate on the motifs $8E_3+12E_2+4E_1$.

The fact that $H_n^{cft}$ act on products of primary fields by a linear
combination of $H_n^{dyn}$ implies that $H_n^{cft}$
commute with each other and with the
Yangian generator \ECa . Then, all the states belonging to an Yangian
multiplet have the same eigenvalue of $H_n^{cft}$ that we evaluated
on a particular (the highest weight) vector.

\listrefs
\end